\documentclass[conference]{IEEEtran}

\usepackage{cite}

\ifCLASSINFOpdf
\else
\fi
\usepackage{amsmath}
\usepackage{algorithmic}

\usepackage{array}

\ifCLASSOPTIONcompsoc
 \usepackage[caption=false,font=normalsize,labelfont=sf,textfont=sf]{subfig}
\else
 \usepackage[caption=false,font=footnotesize]{subfig}
\fi
\usepackage{multirow}
\usepackage{graphicx}

\usepackage{url}

\begin{document}
%
\title{Scaling Data Association for Hypothesis-Oriented MHT}

\author{Michael Motro and Joydeep Ghosh \\%
\IEEEauthorblockA{Department of Electrical and Computer Engineering\\
University of Texas at Austin, USA}
}


\maketitle

\begin{abstract}
Multi-hypothesis tracking is a flexible and intuitive approach to tracking multiple nearby objects. However, the original formulation of its data association step is widely thought to scale poorly with the number of tracked objects. We propose enhancements including handling undetected objects and false measurements without inflating the size of the problem, early stopping during solution calculation, and providing for sparse or gated input. These changes collectively improve the computational time and space requirements of data association so that hundreds or thousands of hypotheses over hundreds of objects may be considered in real time. A multi-sensor simulation demonstrates that scaling up the hypothesis count can significantly improve performance in some applications.
\end{abstract}


%
\IEEEpeerreviewmaketitle

\section{Introduction}
Automated tracking of multiple objects is a valuable task for many security applications as well as robotics \cite{kitti} and biology \cite{bio}. The main challenge of multi-object tracking is considering the uncertainties of object presence, measurement validity, and object-measurement relationships. The standard formulation of multi-object tracking, which considers objects that independently generate single measurements, results in a steadily increasing number of possible object states as well as a correlation between these objects that admits no simple reduction.

Multiple hypothesis tracking (MHT) is one of the original methods to approximate this distribution at each timestep, and has several desirable traits. Its update step is modular with respect to the distribution assumed for each object or the type of single-object tracking used. The number of hypotheses can be chosen to intuitively trade between accuracy and computational cost, and an extremely high number of hypotheses can hypothetically approximate the multi-object state up to arbitrary accuracy.

MHT also has a reputation of being a computationally expensive tracker. This is due first to its consideration of multiple potential matchings for each object at each time, leading to the aforementioned increase in possible object states. The second is data association between objects and measurements, which all multi-object trackers perform implicitly or explicitly in their update steps. Data association for MHT takes many input hypotheses and calculates a similar number of highly likely new hypotheses. Such an algorithm is complex to implement and computationally expensive. MHT primarily sees use when the number of objects is low enough or the problem simple enough that dozens of hypotheses are sufficient \cite{danchick}, or when grouping can be applied \cite{groupinglarge} to separate the problem into smaller ones.

The data association task is equivalent to finding a set of solutions to the bipartite matching, or 2D assignment, problem. Murty’s algorithm is the best known solution, with \cite{miller, crouse, pedersen, lookahead2way} offering modifications to apply it more efficiently to multiple hypothesis tracking. Approximate solutions have also been explored, with random sampling methods in particular gaining traction for their simplicity \cite{randommcmcda, gibbs, randomso}. Track-oriented MHT \cite{blackman, 40years} considers multiple contiguous timesteps at once, thus solving a conceptually similar but algorithmically different data association problem. The original formulation of MHT is often termed hypothesis-oriented MHT to distinguish.

This work revisits the deterministic algorithm for the hypothesis-oriented MHT data association, synthesizes previous improvements, and provides several original improvements. C and Python implementations are available at \url{github.com/motrom/fastmurty}.
Section \ref{background} describes the data association algorithm, while Section III discusses the improvements in implementation. Section IV tests the modified algorithm on simulated data and shows that it outperforms previous versions. It also shows a simple sensor fusion example in which scaling up the number of hypotheses in MHT, as enabled by an optimized implementation, causes a clear improvement in tracking accuracy.

\section{The MHT update algorithm} \label{background}
We start by describing the form of a multiple hypothesis distribution. Each object being tracked is assumed to have a hidden state $x$, and various measurements $z$ are observed at each timepoint. If a certain object is known to exist, and the measurements that correspond to it are known, single-object tracking techniques can be used to estimate the object state with a distribution $p(x)$ and update this estimate according to a measurement model $p(z|x)$. A single \textit{hypothesis} about a multi-object system specifies a set of objects that exist and have estimates $p(x)$. If this hypothesis were certain, and the association between objects in the hypothesis and new measurements were certain, multi-object tracking could be performed by simply storing a vector of distributions $p(x)$ and applying single-object tracking updates to each in parallel. The multiple hypothesis distribution stores a set of $K$ hypotheses, each with a probability of being valid. Hypotheses will generally share many of the same objects, so the multiple hypothesis state can be written compactly as a vector of $M$ single-object distributions, a $K\times M$ binary matrix specifying which objects are present in which hypothesis, and a vector of probabilities for the $K$ hypotheses.

Multiple hypothesis tracking is a Bayesian solution to the tracking problem. Given a prior multiple hypothesis distribution over objects, and a set of $N$ measurements assumed to derive from the standard measurement model, the posterior distribution is also a finite-sized multiple hypothesis distribution. However, in nearly all practical problems the size of this distribution becomes too large to utilize, and an approximation by truncation is performed. A full derivation of the MHT update is not presented here as it has been covered many times in several ways, for instance \cite{40years, williamsrfs}. For any prior hypothesis, there are many possible associations between its tracked objects and the new measurements. We parameterize an association $A$ as a binary matrix in which nonzero elements $A_{ij}$ represent matches between object $i$ and measurements $j$. Objects with no match did not create a measurement, and measurements with no match came from a yet-untracked object or are erroneous. We refer to either as a \textit{miss}. A valid association will not have any one measurement corresponding to two objects, or any one object creating two measurements.
 Each association has a probability based on the likelihood that each measurement is created by each object.
\begin{align} \label{orig_mht_score}
&P(A) = \prod_{i,j : A_{ij}} L_{ij} \prod_{i:\sum_{j} A_{ij}=0} p(i\text{ miss}) \prod_{j:\sum_{i} A_{ij}=0} p(j\text{ miss}) \\
&L_{ij} = \int p_i(x) p(z_j|x) d x \nonumber
\end{align}
Each object and measurement match results in a distinct single-object posterior, so the updated object vector will be of size $MN$. Each association specifies a certain set of prior object and measurement matches and is a hypothesis on the updated object vector. The total probability of each new hypothesis is proportional to the probability of association times the probability of the prior hypothesis it came from.
Thus the MHT update multiplies the number of tracked objects by $N$ and increases the number of hypotheses by a factor of over $N$ factorial\footnote{Runtime and size bounds in this paper assume $M>N$, without loss of generality}. Storing or even iterating over this many hypotheses is infeasible. The practical solution is to locate high-probability associations for each prior hypothesis without searching through all possibilities. As it happens, locating the most likely associations can be related to a common problem in graphical models. Taking a negative log transform of (\ref{orig_mht_score}), and removing a constant term that is the same for all associations, returns:
\begin{align} \label{nll_mht_score}
&NLL(A) = \sum_{i,j : A_{ij}} C_{ij} \\
&C_{ij} = -\log\left( L_{ij}\right) + \log\left( p(i\text{ miss})\right) + \log\left( p(j\text{ miss})\right) \nonumber
\end{align}

The associations with the highest probabilities will have the lowest negative log-likelihoods, so this transforms the problem from constrained maximization to constrained minimization. The term to be minimized is the sum of elements of a matrix $C$. If $M=N$ and object and measurement misses are very unlikely, every row of the data matrix has a match to exactly one column. Finding the minimum-scoring association in this case is equivalent to the well-known 2D assignment problem.

This problem has several approaches, but the successive shortest paths (SSP) approach has been repeatedly shown to perform well for data association \cite{crouse,jvisgood} and in general \cite{assignsurvey}, specifically with the specialized form developed in \cite{jv,jvc,lapCTCS}. The SSP starts with the simple problem of matching a single row to any column, then successively adds rows while maintaining the best assignment across all added rows. This is achieved by minimizing the cost on a reduced matrix $C'=C - u1^T - 1v^T$, where the vectors $u$ and $v$ are set such that $C'=0$ for all previous matches and $C'\geq 0$ on all other elements. Note that adding a constant to a row or column will not change the optimal association because it alters all valid associations by the same amount. Therefore the reduced matrix can be used to find the optimal association including a single new row. Given a single source (row), one or more targets (available columns), and a nonnegative cost matrix ($C'$), the problem at each step becomes the classic shortest path problem as solved by Dijkstra's algorithm. Dijkstra's algorithm as modified for SSP is shown in Figure \ref{SP}. Most of our innovations will be represented as modifications to this core step of the data association algorithm. It is important to note that the cost matrix is not actually modified at each step. Instead the row and column reductions $u$ and $v$ are stored. The method to update $u$ and $v$ at each step is shown in \cite{jv, crouse} and others.

\begin{figure}[ht]
\textbf{input} C, u, v, col2row, newrow, targetcols \\
\textbf{declare} unused\_cols, distance2cols, pathback

\begin{algorithmic}[1]
\STATE row = newrow
\STATE shortestpath = 0
\REPEAT
    \FOR{ col \textbf{in} unused\_cols }
		\STATE path = shortestpath + C[row,col]-u[row]-v[col]
		\IF{ path $<$ distance2cols[col] }
			\STATE distance2cols[col] = path
			\STATE pathback[col] = row
		\ENDIF
	\ENDFOR
	\STATE col = \textbf{argmin}(distance2cols[col] \textbf{for} col \textbf{in} unused\_cols)
	\STATE shortestpath = distance2cols[col]
	\STATE unused\_cols.remove(col)
	\STATE row = col2row[col]
\UNTIL{col in targetcols}
\STATE \COMMENT {use pathback to update col2row}
\STATE \COMMENT {use distance2cols to update u and v}
\end{algorithmic}
\caption{Shortest paths algorithm - as in \cite{jv}}
\label{SP}
\end{figure}

The method to iteratively locate the next best assignments, known as Murty's algorithm, follows a similar strategy of generating solving smaller problems with modified matrices. Given a best association, Murty's algorithm partitions all possible associations into $N$ disjoint sets by marking certain row-column pairs as either required or forbidden. The cost matrix for each partitioned problem is the original matrix with alterations to one row and one column. If the reduced version of the original matrix is available, a single shortest path algorithm again finds the optimal association \cite{miller, networkflowtext}. The runtime of a shortest path algorithm for $M$ objects and $N$ measurements ($N\leq M$) is $O(N^2)$, so the runtimes to obtain the best solution and $K$ best solutions are $O(MN^2)$ and $O(KMN^2)$ respectively.

\section{Optimizations}
\subsection{Missing Objects and Measurements}
\begin{figure}[t]
\centering
\includegraphics[width=3.5in]{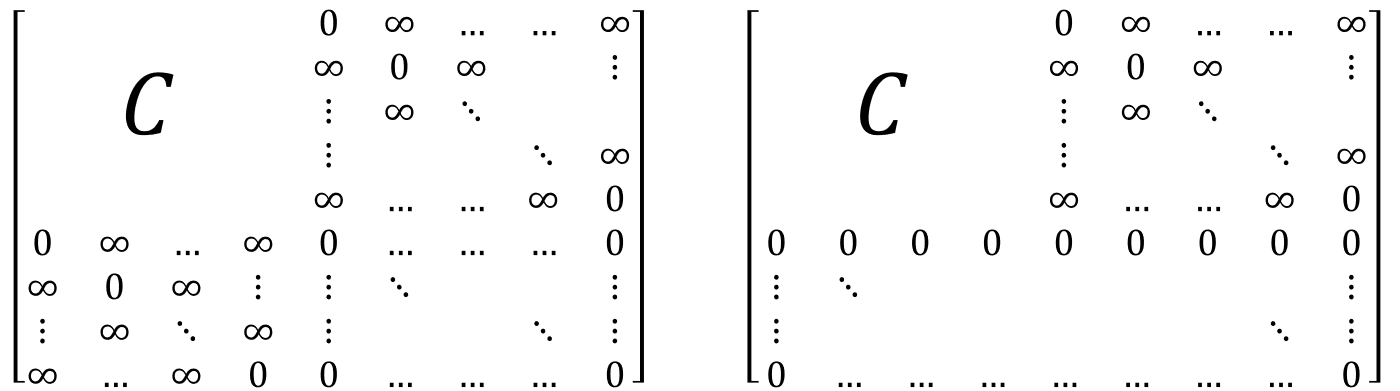}
\caption{Two ways to augment the cost matrix so that missing rows and columns are handled correctly by the original SSP algorithm.}
\label{augmentpic}
\end{figure}

The previously discussed algorithm operates on a square matrix and assumes that every row is matched to a single column and vice versa, $\sum_i A_{ij}=1$. The MHT update must also consider missing rows and columns. Missing rows correspond to objects that were undetected or may not exist, missing columns correspond to measurements from untracked objects or sensor error. The missing case can also be viewed as a special case of a generalized assignment problem \cite{goldenanniversary}. The simplest way to apply the previous algorithm to the miss-enabled problem is to augment the cost matrix with additional rows and columns, as seen in Figure \ref{augmentpic} left. Each augmenting column can only be assigned to one row, and so correctly denotes if that row is missing.

This form is easy to solve but quadruples the size of the cost matrix, thereby increasing the data association runtime. Additionally, note that the bottom right corner of the matrix can fit many possible equivalent matchings. Duplicate solutions must be avoided by modifying the subproblem creation stage of Murty's algorithm. \cite{crouse, augment2, jvc} use the equivalent augmentation shown in Figure \ref{augmentpic} right. The all-zero rows can be ignored during subproblem creation and the SSP algorithm. Additionally, each shortest path step will terminate if it reaches any augmenting column because these columns may each match to an unassigned all-zero row. Thus, SSP with this formulation can handle misses without a notable increase in runtime. However, the step of row and column reduction has not been formulated without a fully augmented matrix. \cite{crouse} for instance perform implicitly-augmented SSP for obtaining the single best association, but resort to the augmented matrix when finding $K$ associations.

We argue that if the reductions are initialized to $0$, a valid solution will be obtained by performing the implicitly-augmented SSP and setting all augmenting row or column reductions to $0$. The simplest way to show this is through the updates of the reductions following a shortest path step. The update procedure is derived in (\cite{assignmenttext} Chapter 4).
\begin{align}
&p_T = \text{shortest path from added row to unmatched column} \\
&p_j = \text{shortest path from added row to column }j \nonumber \\
&v_j = v_j - \max\left(p_T  - p_j, 0\right) \nonumber \\
&u_i = c_{ij'} - v_{j'} \;,\; j' = \text{ updated column matching row }i \nonumber
\end{align}
Note that a column will be set to missing at the end of the SSP if and only if it was never in the shortest path for an added row. Thus a missing column's reduction is never modified from its initial value. If augmenting rows were also added to the SSP after all non-augmenting rows, they would be trivially matched to their corresponding missing column or to an augmenting column. In either case, the shortest path is $0$ and no reductions are altered. Similarly, an augmenting column has an infinite shortest path distance to any non-augmenting rows other than the one it corresponds to. Thus the reduction of an augmenting column for a missing row is only modified when that row is set to missing. At that point, the path to that augmenting column is the shortest path to an unmatched column, so the reduction is again unaltered. Augmenting columns corresponding to matched rows are only reached during the implied addition of augmenting rows. If all reductions are initialized to $0$, then at the end of SSP all augmenting rows and columns will have reduction $0$.

Furthermore, we derive a modification to the shortest paths algorithm for each subproblem that directly considers the structure of the augmented problem and the fact that augmenting rows and columns will have reductions of $0$. This modification is shown in Figure \ref{SPaug}. Note that in the initial SSP, the missing column denoted $j=\theta$ is always a target column, so the method will terminate before searching the augmenting rows and columns. These two modifications reduce the standard data association algorithm to roughly the same runtime as the simpler no-misses algorithm.

\begin{figure}[p]
\begin{algorithmic}[1]
\STATE already\_augmented = \FALSE
\STATE shortestpath = 0
\REPEAT
    \IF {row == $\theta$ \AND \NOT already\_augmented}
        \STATE already\_augmented = \TRUE
        \STATE \COMMENT{ exit augmentation on col}
        \FOR{ col \textbf{in} unused\_cols }
    		 \STATE path = shortestpath - v[col]
    		\IF{ path $<$ distance2cols[col] }
    			\STATE distance2cols[col] = path
    			\STATE pathback[col] = $\theta$
    		\ENDIF
    	\ENDFOR
    	\STATE \COMMENT{ exit augmentation on row}
    	\FOR{row \textbf{in} rows s.t. row2col == $\theta$}
    	    \FOR{ col \textbf{in} unused\_cols }
    		 \STATE path = shortestpath + C[row,col]-u[row]-v[col]
    		\IF{ path $<$ distance2cols[col] }
    			\STATE distance2cols[col] = path
    			\STATE pathback[col] = row
    		\ENDIF
    	\ENDFOR
    	\ENDFOR
    \ELSIF {row $!= \theta$}
        \FOR{ col \textbf{in} unused\_cols }
            \IF{ col==$\theta$ }
                \STATE path = shortestpath - u[row]
            \ELSE
    		    \STATE path = shortestpath + C[row,col]-u[row]-v[col]
    		\ENDIF
    		\IF{ path $<$ distance2cols[col] }
    			\STATE distance2cols[col] = path
    			\STATE pathback[col] = row
    		\ENDIF
    	\ENDFOR
    \ENDIF
	\STATE col = \textbf{argmin}(distance2cols[col] \textbf{for} col \textbf{in} unused\_cols)
	\STATE shortestpath = distance2cols[col]
	\STATE unused\_cols.remove(col)
	\IF{ col == $\theta$ }
	    \STATE \COMMENT{enter augmentation from row}
	    \STATE row = $\theta$
	\ELSE
	    \STATE row = col2row[col]
	    \IF{row == $\theta$}
	        \STATE \COMMENT{enter augmentation from col}
	        \STATE pathback[$\theta$] = col
	        \STATE unused\_cols.remove($\theta$)
	        \STATE col = $\theta$
	    \ENDIF
	\ENDIF
\UNTIL{col in targetcols}
\end{algorithmic}
\caption{Shortest paths algorithm - implicitly handling misses}
\label{SPaug}
\end{figure}

\subsection{Looking Ahead and Early Stopping}
Locating the $K$ best associations on an $M\times N$ cost matrix with Murty's algorithm requires constructing $KN-N+1$ problems. The currently constructed problem with the best solution is located $K$ times. Keeping all problems in an unsorted array incurs $O(K^2N)$ costs while searching for the next best solutions, and keeping the best in a length-$K$ sorted array incurs $O(K^2N)$ for maintaining the order. Partially-ordered data structures offer a faster alternative. \cite{miller, crouse} store problems in a single-ended priority queue, which incurs $O(KN\log (KN))$ cost for adding and deleting problems while maintaining the location of the problem with best remaining solution.

\cite{pedersen} instead use a double-ended priority queue, which has a similar order of runtime but also maintains the worst remaining solution. The size of a double-ended queue can be kept at $K$ instead of $KN-N+1$, with solutions that are worse than the current $K$ are simply discarded. Reducing the number of stored problems is useful from a memory perspective, as most algorithms require $O(M)$ storage for each problem. Storing every constructed problem in a queue results in $O(KMN)$ memory usage, roughly equivalent to copying the cost matrix $K$ times.

However, the single-ended priority queue approach was shown to outspeed the double-ended approach \cite{pedersen} due to a clever optimization it enables. \cite{miller} showed that problems can be placed in the queue before being solved, with an underestimated solution cost. If the lowest-cost problem in the queue is unsolved, it is solved and placed back in the queue. If the lowest-cost problem is solved, then it must have the lowest cost of any solutions potentially in the queue. The final cost of a problem created by Murty's algorithm is the cost of its originating problem, plus the distance returned by a single shortest paths algorithm applied to the reduced cost matrix. A lower bound on the latter cost can be found by looking a single step into the shortest paths algorithm. Assuming the path starts at row $i$:
\begin{equation}
    NLL(A) \geq NLL(\text{orig. }A) + \min_{j : A_{ij}=0} \left( C[i,j]-u[i]-v[j] \right)
\end{equation}

A similar look-ahead estimate is used by \cite{lookahead2way}. Problems with a high enough lower bound will often be left unsolved at the end of the algorithm. This optimization is taken one step further by \cite{miller}. The problems created by Murty's algorithm vary in size, as each has some rows and columns that are fixed to the original problem's solution. If the problem sizes are ordered by their lower bounds, then the largest problems will be the least likely to  be reached before the end of the queue.

Placing unsolved problems in the queue is incompatible with the constraint of storing only $K$ problems. We suggest a different form of optimization for the double-ended priority queue. The $K$th best cost is known at all times, and any problem with a higher cost does not need to be fully solved. Thus the shortest path algorithm should be terminated if the shortest path exceeds the worst currently stored cost. This can be achieved by adding a cost check to the loop exit condition in line 15 of Figure \ref{SP}. We refer to this technique as early stopping. In practice, it has a similar effect to leaving problems unsolved, while requiring nearly no additional computation. It can also be used in conjunction with the look-ahead problem ordering, ensuring that larger problems often have higher costs and are stopped early.

\subsection{Gating}
The previously discussed optimizations do not change the fundamental worst-case runtime of the $K$-best associations algorithm, $O(KMN^2)$. Such worst-case behavior is unavoidable when matches between every row and every column are considered. In practical tracking problems, it is rare for large numbers of objects to all potentially match to large numbers of measurements. In fact, it is common to assume that each object has a small number of matching measurements, even outside the data association step. Typically, an initial similarity check is applied for each measurement and object, with pairs that fail this check being assigned $0$ likelihood. This is termed \textit{gating} \cite{textbook}.

It stands to reason that a data association problem with only a few matches per row/object should be much faster to solve than a fully dense problem. In fact, this possibility was explored by \cite{halfsparse} using an alternative 2D assignment algorithm called the auction algorithm \cite{auction}. Unlike SSP's standard formulation, the auction algorithm can directly handle sparse inputs, and outperformed SSP on highly sparse problems. However, the many optimizations developed for applying Murty's algorithm, such as the ones in this paper and most notably the storage of row and column reductions, have not been replicated for the auction algorithm.

Sparse variants of SSP have been proposed before \cite{lapCTCS} with observed speedup but without a guaranteed reduction in the runtime order. We instead note that there is a variant of Dijkstra's shortest path algorithm that handles sparse input \cite{dijkstrasparse}. This variant uses a single-ended priority queue to efficiently find the nearest column for each iteration, and adds columns to the queue if they can be matched by the current row. The shortest path algorithm for SSP can be modified to fit this formulation as shown in Figure \ref{SPsparse}. No other parts of SSP or Murty's algorithm require major alteration to accommodate gating. If at most $S$ columns can be matched to each row, the runtime of this variant of shortest paths algorithm is $O(MS\log(MS))$, so the overall worst-case runtime of data association reduces to $O(KM^2S\log(MS))$.

\begin{figure}[ht]
\textbf{input} C (sparse), u, v, col2row, newrow, targetcols \\
\textbf{declare} unused\_cols, pathback, pathqueue

\begin{algorithmic}[1]
\STATE row = newrow
\STATE shortestpath = 0
\REPEAT
    \FOR{ col \textbf{in} (unused\_cols $\cap$ C[row]) }
		\STATE path = shortestpath + C[row,col]-u[row]-v[col]
		\STATE pathqueue.push(path, col)
	\ENDFOR
	\REPEAT
	\STATE shortestpath, col = pathqueue.pop()
	\UNTIL{ col \textbf{in} unused\_cols }
	\STATE unused\_cols.remove(col)
	\STATE pathback[col] = row
	\STATE row = col2row[col]
\UNTIL{col in targetcols}
\end{algorithmic}
\caption{Shortest paths algorithm - sparse version}
\label{SPsparse}
\end{figure}

\section{Tests} \label{occlusionsection}
\subsection{Random Dense Problems}
\begin{figure}[t]
\centering
\includegraphics[width=3.1in]{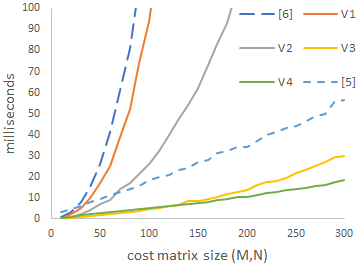}
\caption{Performance of $K$-best association implementations on random matrices with a single input hypothesis.}
\label{randsize2f}
\end{figure}
It is common to assess the computational performance of data association using square matrices with elements sampled from a uniform distribution. We follow \cite{miller} and take a single input hypothesis, 200 output associations, and square matrices of varying size. Four versions of our work are shown, with cumulative improvements as given in Table \ref{versiontable}. The gating for V4 keeps 30 significant elements per row. Our methods are compared to open source code from \cite{miller} and \cite{crouse}. Note that \cite{miller} does not handle missing cases, so the matrices were set such that the top 200 associations would not include missing rows or columns. Our implementation is written in C, while the open source codes are in C and C++ respectively. All were compiled using gcc 7.3.0 with -O3 setting.

\begin{table}[ht]
\centering
\caption{Tested versions of algorithm, with cumulative improvements}
\label{versiontable}
\begin{tabular}{c|l} 
V1 & implicit augmentation for miss \\
V2 & early stopping \\
V3 & look-ahead ordering of problems \\
V4 & gating
\end{tabular}
\end{table}

The results are shown in Figure \ref{randsize2f}. The first axis is the matrix size, and the second is the runtime in milliseconds. The left and right plots are identical except for y-axis scaling, to show the performance of all methods. Our V1 implementation is similar to the code of \cite{crouse} except for implicit handling of misses, so the improvement in runtime can be largely attributed to that modification. V2 is faster than either by an order of magnitude, showing the value of early stopping. However, the look-ahead optimization still gives a substantial advantage to both \cite{miller} and V3. V3 is faster for problems of the tested size but unfortunately does not match the linear scaling property of the single-ended queue with look-ahead. For even larger tracking problems, \cite{miller} outperforms V3 - but in these larger problems the quadratically scaling memory usage of a single-ended queue may become problematic. It is also worth noting that our tested runtime for \cite{miller}'s code is about 40 times faster than that reported by their paper. Decades of hardware improvement have impacted data association performance as much as any algorithm choice.

V4 is slightly slower than V3 for smaller problems, as the gating variant of the shortest paths algorithm is more complex. However, in the long term it scales linearly with input size at a lower rate than \cite{miller}. This stands to reason: we have fixed the number of elements per row, so the number of significant elements in the input matrix is increasing at a linear rather than quadratic rate.

\begin{figure}[ht]
\centering
\includegraphics[width=3.2in]{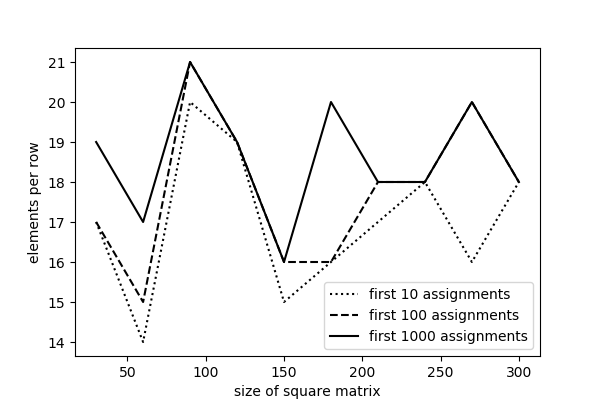}
\caption{Number of gated elements per row that captured all associations on 200 random matrices.}
\label{sparsetest}
\end{figure}

It may be considered inaccurate to use such strong gating on random matrices. For instance, \cite{halfsparse} instead generated sparse matrices for testing. However, our gated version obtained the correct 200 solutions on all tests in this section. This behavior was noted previously by \cite{lapCTCS} and is further examined in Figure \ref{sparsetest}. 200 random matrices were generated at each size, and the most extreme gating that resulted in the correct output on all tests was determined. The gating is inconsistent, but at most 21 elements per row were needed. This is obviously a probabilistic rather than absolute guarantee - the all-ones matrix is a clear example for which gating will eliminate many equally optimal associations. However, most tracking problems are considered to have sparse object-measurement ambiguity, and it is unclear what real problems would have more dense ambuigity than the ones randomly sampled in this test. Hence, we believe a gating-based data association is widely applicable.

\subsection{Comparison to Randomized Data Association}
Random sampling of likely associations has been explored several times in order to find many hypotheses with a low runtime, as well as to solve modified data association problems such as for extended targets. We implemented a gated version of the Gibbs sampling method of \cite{gibbs} in the same environment as our deterministic implementation. Both were applied to random 100x100 problems. Table \ref{tab:random} reports the total likelihood of obtained associations divided by the likelihood of the single best association, as a measure of the usefulness of the obtained associations. For the deterministic algorithm, the total likelihood scales nearly linearly with associations obtained, meaning each association is nearly as likely as the most likely one. The total likelihood of the random algorithm increases at a very low rate, as predominantly unlikely associations are sampled and some associations are sampled several times. On this type of input, deterministic data association clearly outperforms the explored random approach.

\begin{table}[ht]
\caption{Performance of Deterministic and Random Data Association}
\label{tab:random}
\begin{tabular}{|c|c|c|c|} \hline
method & associations computed & likelihood ratio & runtime (ms) \\ \hline
Deterministic & 10 & 9.96 & 0.67 \\
Deterministic & 1000 & 982. & 29.7 \\ \hline
Random & 10 & 2.93 & 0.15 \\
Random & 10000 & 6.82 & 67.5 \\ \hline
\end{tabular}
\end{table}

\subsection{Multiple Input Tests}
The previous tests assume a single input hypothesis, but MHT is in fact a multiple-input-multiple-output (MIMO) problem. We assume that the number of input hypotheses and the number of desired output hypotheses are equal. This does not require solving $K$ separate $K$-best association problems; instead, the problem specified by each hypothesis can be placed in a queue, and the $K$ best solutions will be obtained from the queue in the same manner as with a single hypothesis. The main difference with MIMO data association, from a computational standpoint, is that the best solution for each of the input hypotheses will be obtained before they are placed in the queue. Thus the initial SSP step is run $K$ times instead of once.

\begin{figure}[ht]
\centering
\includegraphics[width=3.1in]{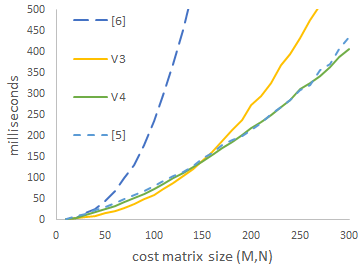}
\caption{Performance of MIMO $K$-best association implementations on random matrices.}
\label{randsize3f}
\end{figure}

Figure \ref{randsize3f} repeats the tests of Figure \ref{randsize2f} with multiple inputs. Two random matrices are used, with the top 200 solutions on the first matrix acting as input hypotheses on the second matrix. Note that the rows for the second matrix correspond to row-column pairs from the first matrix that were included in one of the top solutions. Thus the second matrix is larger than the first, but for any single hypothesis the number of included rows (objects) will be roughly the same. As the codes from \cite{miller, crouse} were not set up for this test, we instead impute their performance by calculating the average runtime of the single SSP, multiplying this by $K-1$, and adding it to their single-input performance. The relative performance of methods is the same as in the single-input results, except that the new algorithms no longer outperform \cite{miller}. It is surprising that \cite{miller} achieves close to linear scaling with matrix size without any sparsity assumption - this is most likely due to the initialization developed by \cite{jv}, which is similar in form to a sparse matrix update. We do not adopt this initialization because it seems to require discretized input and is difficult to adapt for missing rows and columns (which \cite{miller} does not consider). Tracking with many hypotheses at over ten updates a second appears to only be feasible for problems with up to 100 objects - unless more extreme gating is applied. However, this upper limit includes many challenging and relevant tracking problems such as \cite{mot, kitti}.

\subsection{Multi-view sensor fusion}
\begin{figure}[ht]
\centering
\includegraphics[width=3.4in]{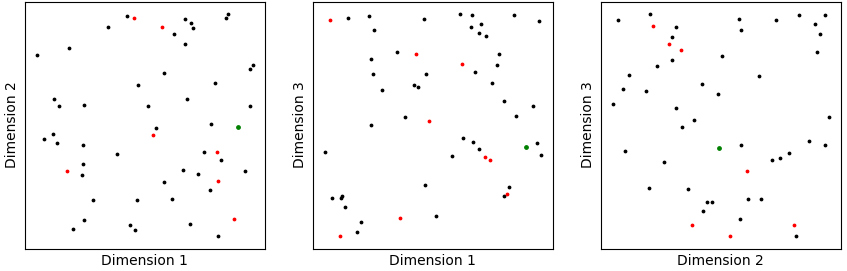}
\caption{Example set of measurements for the multi-view test. Red points represent false positive measurements that are unique to each sensor. The measurements corresponding to one example object have been marked green.}
\label{camexample}
\end{figure}

This section aims to show the value of scaling up the hypothesis count with a simple simulated example. We simulate point targets in 3D space and detect them with three different sensors that each see only two dimensions. Sensor 1 sees dimensions 1 and 2, sensor 2 sees dimensions 1 and 3, and sensor 3 sees dimensions 2 and 3. Figure \ref{camexample} shows an example realization of this problem, from the view of the three sensors. This problem is similar to fusing information from multiple stationary cameras \cite{cameraexample}. While the point targets are not moving,  this problem still violates the Markov property between measurements - sensors 1 and 3 share information that sensor 2 does not access. Thus this problem prohibits simple network flow reductions \cite{netflowSSP}, like many challenging tracking problems. On the other hand, a multi-frame association algorithm would likely perform well if each sensor is accurate.

Our test includes 100 point objects within the unit square, each detected with probability .995. White noise with deviation .001 is applied to each measurement, and false positives occur at a rate of .25 per test. Sensor 1 and 2 are fused with single-input data association, resulting in a set of potential matches between measurement 1 and 2 as well as measurements from either sensor without a match. To reason correctly about the latter without inflating the number of hypotheses, we use the multi-Bernoulli mixture filter \cite{pmbm} and treat the unmatched measurements as objects with a certain probability of existence. We did not prune measurement pairs as there were always less than 300 included within the collected hypotheses. The second data association step matches the first step's pairs and sensor 3's measurements, using the associations output by the first step as input hypotheses. Given the output hypotheses of this step, we follow section IV-A of \cite{pmbm} and report the pairs within the best current hypothesis that have a probability of existence higher than 0.5.

\begin{figure}[ht]
\centering
\includegraphics[width=3.5in]{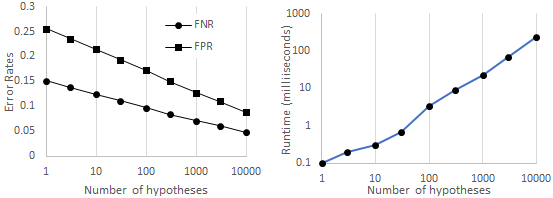}
\caption{Performance of MHT on multi-sensor test. Left: error rates, Right: runtime}
\label{camresults}
\end{figure}

Figure \ref{camresults} shows the performance of the tracker with three metrics. The first is runtime, and the latter show the recovery rate of the correct measurement relationships. False negative rate (FNR) is the ratio of true objects whose set of corresponding measurements was not correctly determined. False positive rate (FPR) is the ratio of reported objects that did not correspond to true objects, or were tracked with an incorrect combination of measurements. Results show improvement in performance as the number of hypotheses increase, at a surprisingly steady log-linear rate. Runtime increases linearly as expected. Even 1000 input and output hypotheses can be handled at over 30 updates per second, and perform roughly twice as well as a single-hypothesis tracker.

\section{Conclusion}
Hypothesis-oriented multiple hypothesis tracking is a classic approach to multiple-object tracking that is hindered by the computational demands of its data association algorithm. We provide three algorithmic improvements to this algorithm. The resulting implementation handles undetected objects and false measurements, its memory scales linearly with the number of objects, measurements, and hypotheses, and its runtime scales linearly (in practice) with the number of objects, measurements, and hypotheses. Other published or publicly available implementations that we are aware of achieve at most one of these properties. With this speed hundreds or thousands of hypotheses can be applied directly to complex sensor fusion problems in real time, without additional modifications or approximations. The tracking community is moving towards open-source libraries for tracking \cite{trackercomponentlibrary, stonesoup, pythonlib}; our implementation is publicly available and will hopefully be a valuable component for future tracking systems.


\section*{Acknowledgment}
This research was supported by Qualcomm as part of the project "Robust and Efficient Multi-Object Tracking for Automotive Applications."




\bibliographystyle{IEEEtran}
\bibliography{IEEEabrv,main}

\end{document}